\documentclass[a4paper,12pt]{iopart}
\pdfoutput=1
\usepackage{iopams,setstack}
\usepackage[utf8x]{inputenc}
\usepackage[pdftex]{graphicx}
\usepackage{amsfonts,latexsym}
\usepackage{amssymb,bm}
\usepackage{calc}
\usepackage{psfrag}
\usepackage{epstopdf}
\usepackage{ifthen}
\usepackage[subnum]{cases}
\usepackage{pgfplotstable}
\usepackage{pgfplots}
\usepackage{tikz}
\usetikzlibrary{intersections,calc,arrows,shapes,patterns,decorations,fadings}

\newcommand{\abs}[1] {\left|#1\right|}
\newcommand{\ccpar}[1] {\left(#1\right)}
\newcommand{\sqpar}[1] {\left[#1\right]}

\newcommand{\ket}[1] {\left| #1 \right>}
\newcommand{\bra}[1] {\left< #1 \right|}
\newcommand{\braket}[2] {\left< #1 \middle| #2 \right>}
\newcommand{\braketop}[3] {\left< #1 \middle| #2 \middle| #3 \right>}

\newcommand{\sgn}[1] {\mathrm{sgn}\ccpar{#1}}

\newcommand{\eqref}[1]{(\ref{#1})}

\definecolor{linkcolour}{rgb}{0,0.2,0.6} 
\definecolor{refcolour}{rgb}{0,0.2,0.6} 

\begin{document}
\title{Analytical solution of electronic transport 
through a benzene molecule using lattice Green's functions}

\author{E. J. C. Dias and N. M. R. Peres}
\address{University of Minho, Physics Department, CFUM, P-4710-057, Braga, Portugal}

\begin{abstract}
Using a Green's function formalism we derive analytical expressions for the 
electronic transmittance through a benzene ring. To motivate the approach we first solve the
resonant level system and then extend the method to the benzene case. These results can be used to
validate numerical methods.
\end{abstract}

\maketitle
\section{Introduction}

The transport through molecules as small as benzene \cite{Reed_1997} or as long as DNA \cite{Slinker_2011}
or carbon nanotubes \cite{Avouris_2002}  
falls into the realm of molecular electronics. This field of research has a long 
history \cite{Ratner_2013}, dating back
to 1956 \cite{Hippel_1956}. As a consequence of this latter paper and of the immediate 
perception of its importance  
the USA Air Force decided, already by that time, 
to fund this type of research \cite{CuevasScheer}. 

According to Cuevas and Scheer \cite{CuevasScheer} molecular electronics is the field of research
devoted to the study of the electronic and thermal transport through individual molecules. The 
objective of this type of research is to produce electronic devices where the building blocks are
single molecules, exploring the quantum mechanical properties that necessarily appear at this small
length scales. This type of devices include: switches, sensors, diodes, transistors, and other
\cite{Ratner_2003}. On the other hand, another advantage of this approach is that it opens the door to the 
discovery of new quantum phenomena. In this sense, one should remember that, historically, 
technological breakthroughs were often a natural consequence of the explorations of new areas, and
not its finality. From a technological point of view, there are several advantages in pursuing the study
of molecular electronics
when compared to the present silicon technology, namely \cite{CuevasScheer,ChoiMody}:
\begin{itemize}
\item Size. Molecules have typical sizes ranging from $1$ to $100$ nm, which allows the construction of
functional nano-structures with subsequent advantages in cost, efficiency and power dissipation.
\item Speed. Estimations show that a good molecular wire could, for example, reduce the transit
time of a traditional transistor, around $10^{-14}$ s, hence obtaining faster operations.
\item New functionalities. As an example, many molecules have several stable geometric structures, 
or isomers, that may have different electronic or optic properties, what can be an advantage for the 
development of some devices, such as switches or sensors.
\item Synthetic tailorability. By changing, for example, the composition or the geometry of a system,
one can obtain specific transport or optical properties suitable for a specific end.
\end{itemize}

\noindent However, we need not only to know which molecules are the best for a certain usage, but also how 
to connect them, namely through unidimensional (1D) nanowires. We shall focus now on the composition of
these nanowires, rather than on how to create them (for that, see, for example [\cite{CuevasScheer}]).

Historically, the carbon-based molecules always played an important role in the field of molecular
electronics, due to its
high versatility. Carbon is the basis of a wide range of structures, since tridimensional, like diamonds 
or graphite, to bidimensional, like graphene, (quasi)unidimensional, like the nanotubes, or even 
(quasi)zero-dimensional, like the fullerenes. In particular, the carbon nanotubes can be used to
connect individual molecules. Also, some experiments on quantum transport through carbon nanotubes
have already been performed \cite{Kasumov_2003}.

\begin{figure}[hb]
\begin{minipage}{0.33\textwidth}
\begin{flushleft}
\centering
\includegraphics{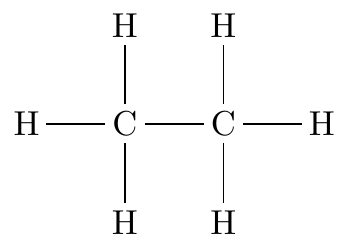}
\label{fig:ethane}
\end{flushleft}
\end{minipage}
\begin{minipage}{0.33\textwidth}
\begin{center}
\centering
\includegraphics{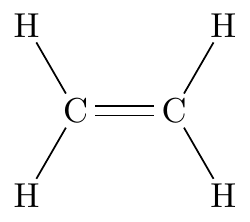}
\label{fig:ethylene}
\end{center}
\end{minipage}
\begin{minipage}{0.33\textwidth}
\begin{flushright}
\centering
\includegraphics{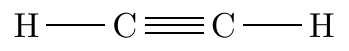}
\label{fig:acetylene}
\end{flushright}
\end{minipage}
\begin{minipage}{0.33\textwidth}
\centering \scriptsize (a) Ethane.
\end{minipage}
\begin{minipage}{0.33\textwidth}
\centering \scriptsize (b) Ethylene.
\end{minipage}
\begin{minipage}{0.33\textwidth}
\centering \scriptsize (c) Acetylene.
\end{minipage}
\caption{This figure shows schematic representations of the simplest of each type of hydrocarbon: 
(a) the ethane, (b) the ethylene and (c) the acetylene. These are, 
respectively, an alkane, an alkene and an alkyne.}
\label{fig:hydrocarbons}
\end{figure}

Another very important class of systems, related to the previous one, are the hydrocarbons. 
The importance of this class
is that such molecules can have different transport properties depending on the degree of hybridization of
the molecular orbitals of its carbon atoms. If the valence electrons are $sp^3$ hybridized, there is a 
tetrahedral arrangement of the bonds in space. This is the case of the alkanes, with the sum formula 
$C_nH_{2n+2}$ [the simplest alkane, the ethane, is presented in figure \ref{fig:hydrocarbons}(a)]. 
Since each valence electron is used for the formation a different (single) chemical bond, 
they are basically
localized, and thus the alkanes are insulating. On the other hand, the valence electrons may also
be $sp^2$ or $sp$ hybridized, forming, respectively, alkenes, with double bonds, and alkynes, with 
triple bonds [the simplest of each kind, the ethylene and the acethylene, are presented in figure 
\ref{fig:hydrocarbons}(b) and (c)]. A particular case of these occurs when double or triple bonds are
alternated with single bonds. These molecules are very stable and give rise to delocalized wave functions,
which makes them good conductors.

\begin{figure}[htbp]
\begin{minipage}{0.6\textwidth}
\begin{flushleft}
\centering
\includegraphics{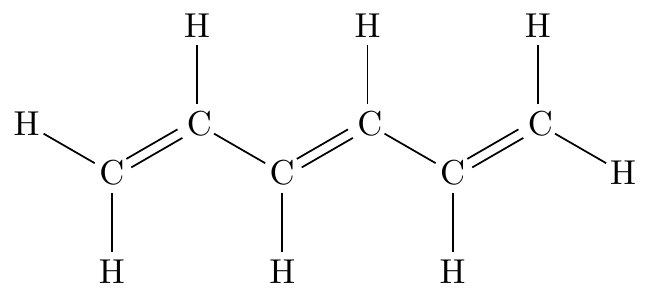}
\label{fig:hexatriene}
\end{flushleft}
\end{minipage}
\begin{minipage}{0.4\textwidth}
\begin{flushright}
\centering
\includegraphics{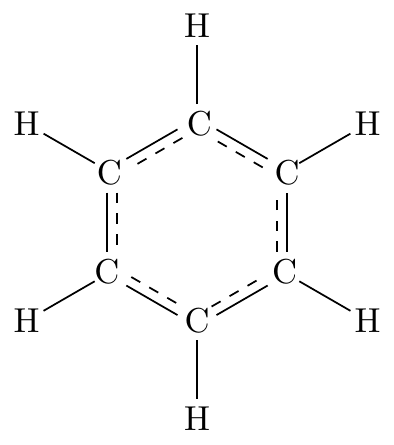}
\label{fig:benzene}
\end{flushright}
\end{minipage}
\begin{minipage}{0.6\textwidth}
\centering \scriptsize (a) Hexatriene.
\end{minipage}
\begin{minipage}{0.4\textwidth}
\centering \scriptsize (b) Benzene.
\end{minipage}
\caption{This figure shows the schematic representation of two different molecules with six carbon atoms each: 
(a) a hexatriene, with zigzag structure, and (b) a benzene, 
with a cyclic structure. The dashed lines on the benzene indicate fully delocalized double bonds.}
\label{fig:polyenes}
\end{figure}

Another advantage of these organic molecules is that they can assume different geometries. For example,
in figure \ref{fig:polyenes} are presented two polyenes with six carbon atoms each. Due
to their atoms' $sp^2$ hybridization, they tend to make 120$^{\rm{o}}$ angles between bonds. When the
angles are built to alternate sides, the polyene has a zigzag shape 
[hexatriene, figure \ref{fig:polyenes}(a)]. 
However, when all angles are built to the same side, the polyene has a cyclic shape [benzene, 
figure \ref{fig:polyenes}(b)]. In the particular case of the benzene, there is a fully delocalized
double bond, which makes this molecule a good conductor. This is the motivation of our study
of the transport through a benzene ring.

The main goal of this paper is to provide a theoretical introduction to molecular electronics,
studying the electronic transport through a benzene ring, using the formalism of Green's
functions. The transport through quantum rings has been considered in the literature before
\cite{Dias_2013}, but the approach used cannot be easily generalized to more complex problems.
On the other hand, the Green's function approach is the method of choice when the complexity 
of the problem increases. The use of the  Green's function method to describe impurity states in a 
tight-binding models  has also been considered \cite{Adame2012}. 
The effect of phonons on transport across a Benzene molecule has also been considered using Green's
function methods \cite{Knap}.

The paper is written with the graduate students in mind. The method of Green's
function is often considered a difficult one, casting a way many students from its study. 
Here we show, by two examples
of moderate complexity, that this is not the case. These two problems are solved analytically, which,
to our best knowledge has not been solved in the literature before using the given method. 
Indeed, although, there are 
many books on quantum transport, none give a detailed discussion of problems of moderate complexity.
Typically,  they solve the one-dimension tight-binding model \cite{Datta1,Datta2}.
The same happens with review articles \cite{Pecchia}.	
We thus consider that the solution of this problem is a valuable addition to the literature.

\section{Formalism and the resonant level system}

The resonant level system is the simplest example of molecular electronics displaying non-trivial
transport properties (this model also describes the transport through a single quantum level
\cite{Jauho}). 
We introduce the formalism used in the next section for describing
electronic transport through a benzene molecule with a {\it hands-on} philosophy by solving the 
resonant level system using Green's functions. This approach will benefit the students wanting
to have a working knowledge of the method.

To explain the basics of the Green's functions formalism, 
let us consider a problem that can be described by a Hamiltonian
written in the form
\begin{equation}
 H=H_0 + V,
\end{equation} 
\noindent such that
\begin{equation}
 H_0 \ket{\psi_0} = E \ket{\psi_0}
 \label{eq:schro1}
\end{equation} 
and
\begin{equation}
 H \ket{\psi} = E \ket{\psi}.
 \label{eq:schro2}
\end{equation} 

Let us also consider that $\ket{\psi_0}$ and $\ket{\psi}$ are such that they can be related by
\begin{equation}
 \ket{\psi} = \ket{\psi_0} + \ket{\psi_S},
\end{equation} 
\noindent where $\ket{\psi_S}$ represents the scattered component of the wave. From the previous
equations, one can easily find the Lippmann-Schwinger equation, of the form
\begin{equation}
 \ket{\psi} = \ket{\psi_0} + G_0 \cdot V \ket{\psi},
 \label{eq:green3:lippmann1}
\end{equation} 
\noindent where we define the free Green's function operator as \cite{Economou}
\begin{equation}
G_{0} \equiv \frac{1}{E-H_0+i\epsilon}.
\label{eq:green3:greenfunction1}
\end{equation} 
\noindent This definition is obtained naturally except for the term $i \epsilon$, where $\epsilon$ is
a infinitesimal positive real number. This term is added to the definition to guarantee that this 
operator describes a wave propagating from the left to the right.

Furthermore, one can replace the term $\ket{\psi}$ on 
right hand side of the Lippmann-Schwinger equation (\ref{eq:green3:lippmann1})  with itself  
obtaining by iteration 
\begin{equation}
 \ket{\psi} = \ket{\psi_0} + G \cdot V \ket{\psi_0},
 \label{eq:green3:lippmann2}
\end{equation} 
\noindent where $G$ is the full Green's function operator, that can be defined as
\begin{equation}
G \equiv \frac{1}{1-G_0 \cdot V} \cdot G_0.
\label{eq:green3:greenfunction2}
\end{equation} 
The advantage of Eq. (\ref{eq:green3:lippmann2}) is that only the known wave function 
$\ket{\psi_0}$ enters in it. Recalling the explicit definition of $G_0$, 
given in \eqref{eq:green3:greenfunction1},
the previous equation can be simplified to
\begin{equation}
G \equiv \frac{1}{E-H_0-V+i\epsilon},
\label{eq:green3:greenfunction3}
\end{equation} 
\noindent which results in a definition analogue to that of $G_0$ except that 
now the full Hamiltonian enters in the definition, and with $\epsilon$ having the same 
meaning as before.
However, neither of the two previous definitions of $G$ are very useful, since both of them involve
inversions of operators, which can be difficult. Fortunately, the Green's function 
operator can yet be written in a
third form, through a Taylor expansion of the term 
$\ccpar{1-G_0 \cdot V}^{-1}$ in \eqref{eq:green3:greenfunction2}.
In doing so, one reaches Dyson's equation
\begin{equation}
 G = G_0 + G_0 \cdot V \cdot G.
\end{equation} 
\noindent Its usefulness will be proven further in the text.

The key aspect of this method is to use the previous relations in order to obtain, in the end, an expression of the form
\begin{numcases}{\braket{n}{\psi}=\label{eq:green3:sistema1}}
C \ccpar{ e^{i k a n} + r e^{-i k a n}}, & $n<0$,  \label{eq:green3:sistema1_r} \\
C \ccpar{ \tau e^{i k a n}},  & $n>0$. \label{eq:green3:sistema1_tau}
\end{numcases}

\noindent where $C$ is some constant and $\ket{n}$ is a position state (we note that we have in mind
a description of our systems by tight-binding Hamiltonians; therefore $\ket{n}$ is a Wannier state). 
From an expression of this
form, it is possible to obtain $r$ and $\tau$, which are respectively the reflection and transmission
amplitudes, such that the reflectance and transmittance through some defect, quantum dot, or molecule
are given by
\begin{equation}
 {\cal R} = \abs{r}^2
\end{equation} 
and
\begin{equation}
 {\cal T} = \abs{\tau}^2.
\end{equation} 

\begin{figure}[htbp]
\centering
\includegraphics{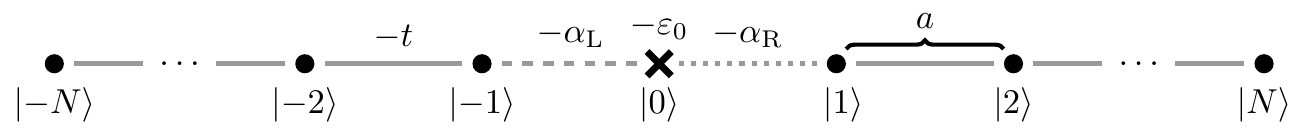}
\caption{Schematic representation of the one level resonant system considered. This system is composed
of two 1D chains with $N$ atoms each, represented by circles, and a defect in the middle of them, 
represented by a cross. In this figure are also identified the hopping energies between different
atoms ($-t$, $-\alpha_{\mathrm{L}}$ or $-\alpha_{\mathrm{R}}$, accordingly), the site energy on the defect ($-\varepsilon_0$)
and the distance between two neighbours ($a$).}
\label{fig:OneLevelResonant}
\end{figure}

We now proceed to the study of the resonant level system. This problem will be useful to 
illustrate the previously described method in a situation where the solution is non-trivial
(in general, textbooks tend to describe only trivial examples which produce some discomfort on the 
students). To start with,
let us define the system we ought to study. This is composed of two semi-finite 1D chains, with $N$ atoms
each, and a different atom in between them -- which we shall call the defect. The distance 
between each atom and its nearest two neighbours is $a$. In both 1D chains, the hopping energy between 
two first neighbours is $-t$, whereas the hopping energy between higher order neighbours is zero. 
However, due to the presence of the defect, we shall consider that, on the one hand,
the hopping energies around the defect are different from $-t$ and are different in both sides of it --
being $-\alpha_{\mathrm{L}}$ to the left and $-\alpha_{\mathrm{R}}$ to the right --, and, 
on the other hand, there is an 
on-site energy on the position of the defect ($n=0$), with value $-\varepsilon_0$. 
In figure \ref{fig:OneLevelResonant}
there is a schematic representation of the described system.

From the previous description, it becomes clear that the Hamiltonian which describes this system
can be written as
\begin{equation}
 H = H_{\mathrm{L}} + H_{\mathrm{R}} + H_{\mathrm{C}} + V \equiv H_0 + V,
\end{equation} 
\noindent where we shall define
\begin{equation}
H_{\mathrm{L}} = -t \sum_{n=-N}^{-2} \Big( \ket{n}\bra{n+1} + \ket{n+1}\bra{n} \Big),
\label{eq:green3:HL}
\end{equation} 
\begin{equation}
H_{\mathrm{C}} = -\varepsilon_0 \ket{0}\bra{0} ,
\label{eq:green3:HC}
\end{equation} 
\begin{equation}
H_{\mathrm{R}} = -t \sum_{n=1}^{N-1} \Big( \ket{n}\bra{n+1} + \ket{n+1}\bra{n} \Big)
\label{eq:green3:HR}
\end{equation} 
\noindent and
\begin{equation}
V = -\alpha_{\mathrm{L}} \Big( \ket{-1}\bra{0} + \ket{0}\bra{-1} \Big) -\alpha_{\mathrm{R}} \Big( \ket{0}\bra{1} + \ket{1}\bra{0} \Big).
 \label{eq:green3:V}
\end{equation} 

We assume that in the beginning of times (say $t=-\infty$) the two semi-infinite chains 
(often called  leads
in  molecular electronics terminology) are decoupled from the central atom.
In this case a flux of electrons incoming from the left of the defect ($n<0$) cannot 
continue to the $n=0$ position being reflected in the end of the semi-infinite lead. 
This means that at $t=-\infty$ the unperturbed wave function must read
\begin{numcases}{\ket{\psi_0}=\label{eq:green3:psi0}}
\ket{\psi_{\mathrm{L}}}, & $n<0$,  \label{eq:green3:psi0L} \\
0,  & $n>0$, \label{eq:green3:psi0R}
\end{numcases} 
\noindent where $\ket{\psi_{\mathrm{L}}}$ is such that
\begin{equation}
 H_{\mathrm{L}} \ket{\psi_{\mathrm{L}}} = E \ket{\psi_{\mathrm{L}}},
 \label{eq:schropsiL}
\end{equation} 
where $\ket{\psi_{\mathrm{L}}}$ plays the role of $\ket{\psi_0}$ in the formalism described above.
\noindent Hence, to apply the previously described method, we must find $\ket{\psi_{\mathrm{L}}}$. 
It can be written, in general, in a basis composed by every position state available, that is,
\begin{equation}
 \ket{\psi_{\mathrm{L}}} = \sum_{m=-N}^{-1} c(m) \ket{m},
\end{equation} 
\noindent where we do not know yet the explicit form of the coefficients $c(m)$. However, we 
can use this result, as well as the definition of $H_{\mathrm{L}}$ given in \eqref{eq:green3:HL},
and replace them in \eqref{eq:schropsiL}. Multiplying  both sides of the 
resulting equation with the position state $\bra{l}$ ($l=-N,\ldots,-1$), one gets that
(called the tight-binding equations)
\begin{numcases}{}
 -t c(-N+1) = E c(-N), & $l=-N$, \\
 -t \sqpar{c(l+1)+c(l-1)} = E c(l), & $-N<l<-1$, \label{eq:green3:schrodinger1}\\
 -t c(-2) = E c(-1), & $l=-1$.
\end{numcases}
\noindent It should be clear that \eqref{eq:green3:schrodinger1} 
becomes general if we introduce the condition
\begin{equation}
 c(-N-1) = c(0) = 0,
\end{equation} 
where it is assumed that $N$ will tend to infinite in the of the calculation.
On the other hand, we can consider that the flux must be composed by the sum of two plane waves, 
one travelling forward and the other travelling backward, which means that the coefficients $c(m)$ 
must have the form
\begin{equation}
 c(m) = C e^{i kam} + D e^{-i kam},
\end{equation} 
\noindent where $C$ and $D$ are constants and $k$ is the wave vector of the plane waves. From these
two conclusions, one finds that
\begin{equation}
 c(m) = C \sin \ccpar{k a m}
 \label{eq:c(m)}
\end{equation} 
and, also,
\begin{equation}
 k \equiv k_n = \frac{n \pi}{a (N+1)},
 \label{eq:k1D}
\end{equation} 
\noindent where $n$ is some integer value between $1$ and $N$. Through further normalization of
the wave function, one finally obtains
\begin{equation}
 \ket{\psi_{\mathrm{L}}} = \sqrt{\frac{2}{N+1}} \sum_{m=-N}^{-1} \sin \ccpar{k a m} \ket{m},
 \label{eq:green3:psiL}
\end{equation}
and also, from the \eqref{eq:green3:psi0},
\begin{equation}
 \braket{n}{\psi_0} = \braket{n}{\psi_{\mathrm{L}}} = \sqrt{\frac{2}{N+1}} \sin \ccpar{k a n},
 \label{eq:green3:npsi0}
\end{equation} 
\noindent if $n<0$, and zero otherwise.
Moreover, it is also possible to find, from \eqref{eq:green3:schrodinger1} and \eqref{eq:c(m)}, 
that the allowed stationary energies of this problem are given by
\begin{equation}
 E(k)=-2t \cos\ccpar{ka}.
 \label{eq:green3:EL}
\end{equation} 

In analogue way,
it is also possible to show that for the 1D chain to the right of the defect
the eigenstates of $H_{\mathrm{R}}$ are given by
\begin{equation}
 \ket{\psi_{\mathrm{R}}} = \sqrt{\frac{2}{N+1}} \sum_{m=1}^{N} \sin \ccpar{k a m} \ket{m},
 \label{eq:green3:psiR}
\end{equation}
and the allowed stationary energies are equally given by \eqref{eq:green3:psiL}. For this
case, $k$ is defined as before, see \eqref{eq:k1D}. This results will be useful later.

Now that  the coefficients $\braket{n}{\psi_0}$ have been found, we can use the Lippmann-Schwinger
equation, as given in \eqref{eq:green3:lippmann2}, and multiply this equation by the position
state $\bra{n}$ on both sides, obtaining [after recalling the definition of $V$ in \eqref{eq:green3:V}]
\begin{numcases}{\braket{n}{\psi}= \label{eq:green3:psi03}}
  \sqrt{\frac{2}{N+1}} \sin \ccpar{k a n} -\alpha_{\mathrm{L}} \braketop{n}{G}{0}\sqrt{\frac{2}{N+1}} \sin \ccpar{-k a}, & $n<0$, \\
  -\alpha_{\mathrm{L}} \braketop{n}{G}{0}\sqrt{\frac{2}{N+1}} \sin \ccpar{-k a}, & $n>0$.
\end{numcases} 

In order to write an equation of the form of \eqref{eq:green3:sistema1}, we must now find
the matrix element $\braketop{n}{G}{0}$ of the operator $G$. Using the Dyson equation,
and considering that, if $n \neq 0$, $\braketop{n}{G_0}{m}$ is only non-zero if
$n$ and $m$ are both positive or both negative\cite{footnote1} one gets
\begin{numcases}{\braketop{n}{G}{0}= \label{eq:green3:G3}}
 -\alpha_{\mathrm{L}} \braketop{n}{G_0}{-1}\braketop{0}{G}{0},  & $n<0$, \\
 -\alpha_{\mathrm{R}} \braketop{n}{G_0}{1}\braketop{0}{G}{0},  & $n>0$.
\end{numcases}
\noindent This means it is now necessary to find the element $\braketop{0}{G}{0}$. This can be done
by solving the system
\begin{numcases}{}
  \braketop{0}{G}{0} = \braketop{0}{G_0}{0} \sqpar{1 -\alpha_{\mathrm{L}} \braketop{-1}{G}{0} - \alpha_{\mathrm{R}} \braketop{1}{G}{0}}, \\
  \braketop{-1}{G}{0} = -\alpha_{\mathrm{L}} \braketop{-1}{G_0}{-1}\braketop{0}{G}{0}, \\
  \braketop{1}{G}{0} = -\alpha_{\mathrm{R}} \braketop{1}{G_0}{1}\braketop{0}{G}{0}, 
\end{numcases}
\noindent whose equations were found using the Dyson's equation over again. By doing so, and defining 
$G_0(n,m) \equiv \braketop{n}{G_0}{m}$ and also $G(n,m) \equiv \braketop{n}{G}{m}$, we get
\begin{equation}
 G(0,0) = \frac{G_0(0,0)}{1-G_0(0,0) \sqpar{\alpha_{\mathrm{L}}^2 G_0(-1,-1) + \alpha_{\mathrm{R}}^2 G_0(1,1)}}.
 \label{eq:green3:G00}
\end{equation} 

To close the problem, it is now necessary to find the matrix elements of the operator $G_0$. That
calculation is presented in \ref{sec:anexoG03}, where it was found the result
\begin{numcases}{ G_0(n,m) =\label{eq:green3:G0nm}}
 \frac{1}{E+\varepsilon_0}, & $n=m=0$, \\
 \frac{i}{2t\sqrt{1-\ccpar{\frac{E}{2t}}^2}} \ccpar{e^{i k a \abs{n+m}} - e^{i k a \abs{n-m}}}, & $n \times m>0$, \\
 0, & otherwise.
\end{numcases} 

Using this result, and defining now the dimensionless variables $Z \equiv {E}/{2t}$, $X \equiv {\varepsilon_0}/{t}$,
 $Y_{\mathrm{L}} \equiv {\alpha_{\mathrm{L}}}/{t}$ and $Y_{\mathrm{R}} \equiv {\alpha_{\mathrm{R}}}/{t}$,
\noindent \eqref{eq:green3:psi03} simplifies to
\begin{scriptsize}
\begin{numcases}{\braket{n}{\psi}= \label{eq:green3:psi06}}
  \sqrt{\frac{2}{N+1}} \sqpar{ \frac{e^{i k a n}}{2i} + \sqpar{\frac{i Y_{\mathrm{L}}^2 \sqrt{1-Z^2}}{Z+\frac{X}{2}+\ccpar{\frac{Y_{\mathrm{L}}^2+Y_{\mathrm{R}}^2}{2}}\ccpar{-Z+i\sqrt{1-Z^2}}}-1} \frac{e^{-i k a n}}{2i}}, & $n<0$, \\
  \sqrt{\frac{2}{N+1}} \sqpar{\frac{i Y_{\mathrm{L}} Y_{\mathrm{R}} \sqrt{1-Z^2}}{Z+\frac{X}{2}+\ccpar{\frac{Y_{\mathrm{L}}^2+Y_{\mathrm{R}}^2}{2}}\ccpar{-Z+i\sqrt{1-Z^2}}}} \frac{e^{i k a n}}{2 i}, & $n>0$.
\end{numcases} 
\end{scriptsize}

\noindent where the sine functions were written in the Euler notation, $\sin x = \ccpar{e^{i x}-e^{-i x}}/\ccpar{2i}$.
Now, this result is in the wanted form, whereby comparing it to \eqref{eq:green3:sistema1},
one finds that
\begin{numcases}{\label{eq:green3:coefs}}
  r=\frac{i Y_{\mathrm{L}}^2 \sqrt{1-Z^2}}{Z+\frac{X}{2}+\ccpar{\frac{Y_{\mathrm{L}}^2+Y_{\mathrm{R}}^2}{2}}\ccpar{-Z+i\sqrt{1-Z^2}}}-1, \\
  \tau=\frac{i Y_{\mathrm{L}} Y_{\mathrm{R}} \sqrt{1-Z^2}}{Z+\frac{X}{2}+\ccpar{\frac{Y_{\mathrm{L}}^2+Y_{\mathrm{R}}^2}{2}}\ccpar{-Z+i\sqrt{1-Z^2}}},
\end{numcases} 
\noindent and, also,
\begin{numcases}{\label{eq:green3:coefs2}}
  \mathcal{R}=\frac{ \ccpar{\frac{Y_{\mathrm{L}}^2-Y_{\mathrm{R}}^2}{2}}^2 + \sqpar{Z\ccpar{1-Y_{\mathrm{L}}^2} + \frac{X}{2}} \sqpar{Z\ccpar{1-Y_{\mathrm{R}}^2} + \frac{X}{2}}  }{\ccpar{\frac{Y_{\mathrm{L}}^2+Y_{\mathrm{R}}^2}{2}}^2 + \ccpar{Z+\frac{X}{2}}\sqpar{Z\ccpar{1-Y_{\mathrm{L}}^2-Y_{\mathrm{R}}^2} + \frac{X}{2}}}, \\
  \mathcal{T}=\frac{Y_{\mathrm{L}}^2 Y_{\mathrm{R}}^2 \ccpar{1-Z^2}}{\ccpar{\frac{Y_{\mathrm{L}}^2+Y_{\mathrm{R}}^2}{2}}^2 + \ccpar{Z+\frac{X}{2}}\sqpar{Z\ccpar{1-Y_{\mathrm{L}}^2-Y_{\mathrm{R}}^2} + \frac{X}{2}}}.
\end{numcases} 

\begin{figure}[htbp]
\centering
\includegraphics[width=\textwidth]{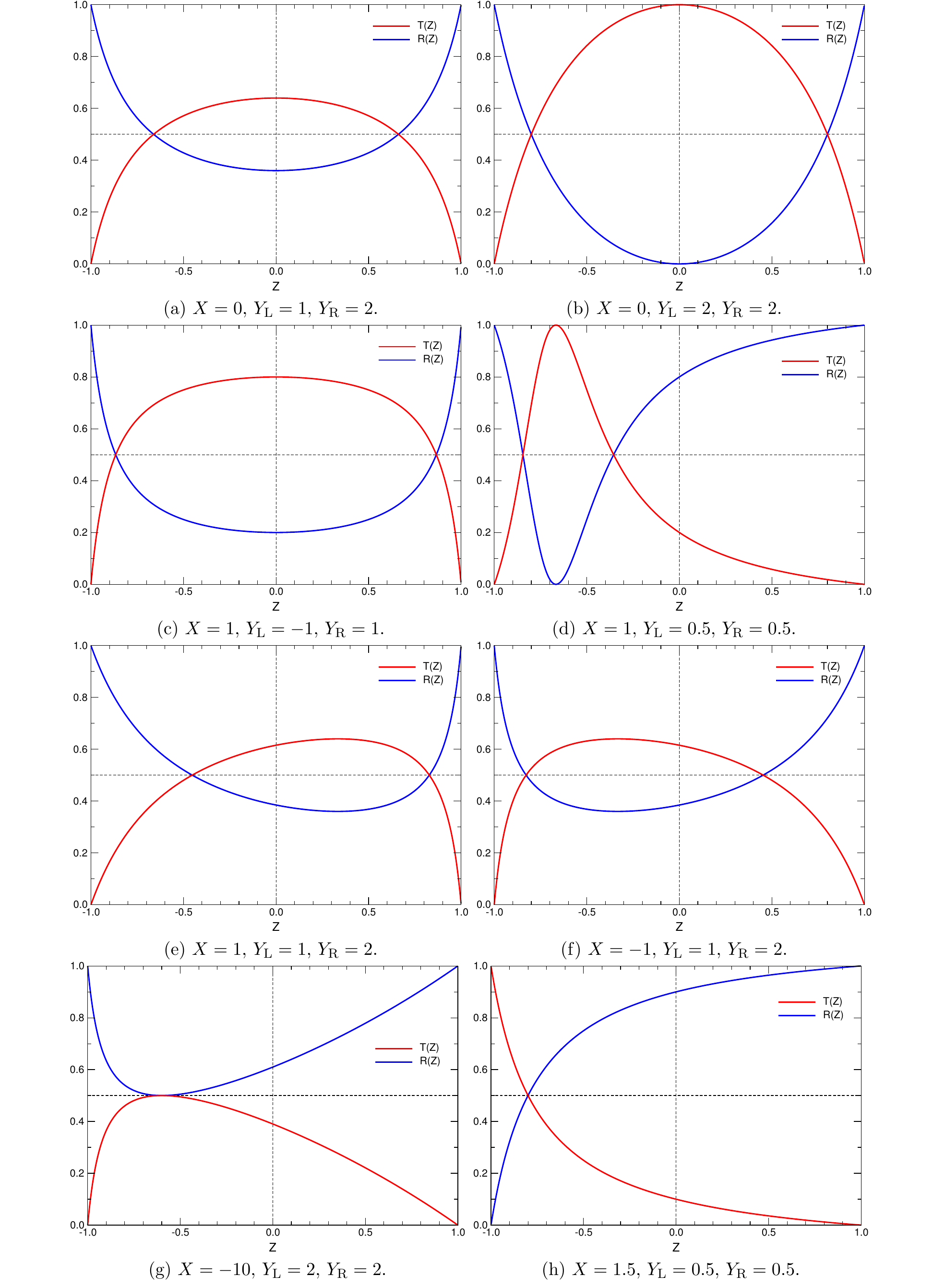}
\caption{Graphic representations of the transmittance and reflectance for the resonant level system, as a 
function of energy, for different values of the parameters $X$, $Y_{\mathrm{L}}$ and $Y_{\mathrm{R}}$.}
\label{fig:TR_resonant_level}
\end{figure}

\noindent In particular, as expected,
\begin{equation}
 \mathcal{R} + \mathcal{T} = 1.
\end{equation} 
It is worth stressing that the formalism allowed us to obtained analytical equations for 
both $\mathcal{R}$ and $\mathcal{T}$.
In figure \ref{fig:TR_resonant_level}, we present some graphical representations of $\mathcal{R}$ and $\mathcal{T}$
as function of the energy $Z$, for several combinations of the parameters
$X$, $Y_{\mathrm{L}}$ and $Y_{\mathrm{R}}$. We leave the discussion of these results to the 
final section of the paper (which can either be read immediately or afterwards).

\section{Transmittance through a benzene ring}

We now turn to the central part of this work, where using the formalism above we compute the 
electronic transmittance through a benzene ring. Also the local density of states at the different 
carbon atoms of the ring is given.

\begin{figure}[htbp]
\centering
\includegraphics{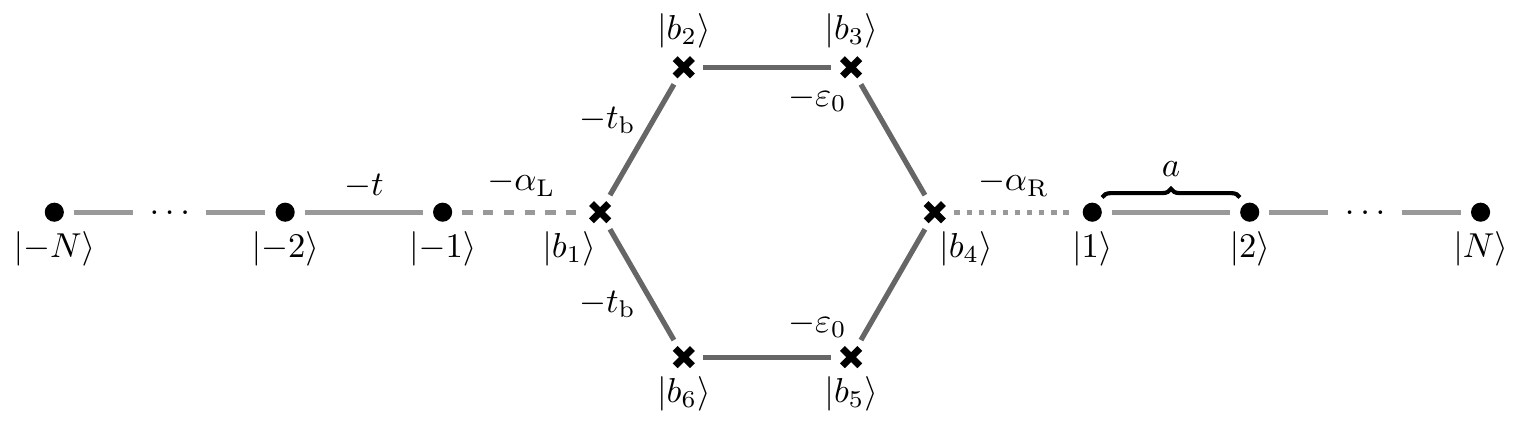}
\caption{Schematic representation of the benzene system considered. This system is composed
of two 1D chains with $N$ atoms each, represented by circles, and a benzene molecule in the middle 
of them. In this figure are also identified the hopping energies between different
atoms ($-t$, $-t_{\mathrm{b}}$, $-\alpha_{\mathrm{L}}$ or $-\alpha_{\mathrm{R}}$, accordingly), the site energy on the 
atoms of the benzene ($-\varepsilon_0$) and the distance between two neighbours ($a$).}
\label{fig:1DChainBenzene}
\end{figure}

In order to do so, we start by defining the system we shall study. It is quite similar to the
level resonant system, but now the defect in not a different atom from the rest of the chain, but a benzene molecule connecting
the two 1D chains, as represented in figure \ref{fig:1DChainBenzene}. These chains are equivalent to 
the ones presented before, and the hopping energies both on the chains and around the defect are equal 
as well. However, now the {\it defect} is composed by a benzene molecule, with all the carbon atoms
placed
at a distance $a$ from its nearest neighbours, and the hopping energy between these atoms 
being $-t_{\mathrm{b}}$.
We also consider on-site energies in the positions of every atom of the benzene molecule, all of them
with the value $-\varepsilon_0$. This makes the model more realistic, since there is a priori no particular
reason for the on-site energy in the leads be equal to that in the benzene.

The Hamiltonian of the given system can be written as
\begin{eqnarray}
 H =H_{\mathrm{L}}+H_{\mathrm{C}}+H_{\mathrm{R}}+V \equiv H_0+V,
 \label{eq:green4:H}
\end{eqnarray} 

\noindent with
\begin{equation}
H_{\mathrm{L}} = -t \sum_{n=-N}^{-2} \Big( \ket{n}\bra{n+1} + \ket{n+1}\bra{n} \Big),
\label{eq:green4:HL}
\end{equation} 
\begin{eqnarray}
H_{\mathrm{C}} = \sum_{i=1}^{6} \Big[ -t_{\mathrm{b}} \Big( \ket{b_{i}}\bra{b_{i+1}} + \ket{b_{i+1}}\bra{b_{i}} \Big) -\varepsilon_0 \ket{b_i}\bra{b_i} \Big],
\label{eq:green4:HC}
\end{eqnarray} 
\begin{equation}
H_{\mathrm{R}} = -t \sum_{n=1}^{N-1} \Big( \ket{n}\bra{n+1} + \ket{n+1}\bra{n} \Big)
\label{eq:green4:HR}
\end{equation} 
\noindent and
\begin{equation}
V = -\alpha_{\mathrm{L}} \Big( \ket{-1}\bra{b_1} + \ket{b_1}\bra{-1} \Big) -\alpha_{\mathrm{R}} \Big( \ket{b_4}\bra{1} + \ket{1}\bra{b_4} \Big),
 \label{eq:green4:V}
\end{equation} 

\noindent where, due to the cyclic structure of the benzene molecule, $\ket{b_{i+6}}=\ket{b_{i}}$.

Considering now, as before, a flux incoming from the left of the defect, then \eqref{eq:green3:psi0}
is still valid. On the other hand, the 1D chains surrounding the benzene molecule are similar as the ones
considered in the previous example, which means that \eqref{eq:k1D}, \eqref{eq:green3:psiL},
\eqref{eq:green3:EL} and \eqref{eq:green3:psiR} are also still valid. As a consequence, the same is true
for \eqref{eq:green3:npsi0}, for $n<0$, and zero otherwise. With this result, we can use the Lippmann-Schwinger
equation, as given in \eqref{eq:green3:lippmann2}, and proceed the same way as before, 
obtaining
\begin{numcases}{\braket{n}{\psi}= \label{eq:green4:psi03}}
  \sqrt{\frac{2}{N+1}} \sin \ccpar{k a n} -\alpha_{\mathrm{L}} \braketop{n}{G}{b_1}\sqrt{\frac{2}{N+1}} \sin \ccpar{-k a}, & $n<0$, \\
  -\alpha_{\mathrm{L}} \braketop{n}{G}{b_1}\sqrt{\frac{2}{N+1}} \sin \ccpar{-k a}, & $n>0$.
\end{numcases} 

We must now find the matrix element $\braketop{n}{G}{b_1}$ of the operator $G$. Using the Dyson equation,
and considering that, in the cases in which $n \neq b_{1,...,6}$, $\braketop{n}{G_0}{m}$ is only non-zero if
$n$ and $m$ are both positive or both negative, we get
\begin{numcases}{\braketop{n}{G}{b_1}= \label{eq:green4:G3}}
 -\alpha_{\mathrm{L}} \braketop{n}{G_0}{-1}\braketop{b_1}{G}{b_1},  & $n<0$, \\
 -\alpha_{\mathrm{R}} \braketop{n}{G_0}{1}\braketop{b_4}{G}{b_1},  & $n>0$,
\end{numcases}
\noindent We need now to calculate the matrix elements $\braketop{n}{G_0}{-1}$, $\braketop{n}{G_0}{1}$, $\braketop{b_1}{G}{b_1}$ 
and $\braketop{b_4}{G}{b_1}$. Due to the higher complexity of this case, we start now by the calculation of the matrix elements 
of $G_0$ instead of the ones of $G$, as done in the previous example. To do so, we use the result obtained in \ref{sec:anexoG04},
\begin{numcases}{ G_0(n,m) =\label{eq:green4:G0nm}}
 \frac{i}{2t\sqrt{1-\ccpar{\frac{E}{2t}}^2}} \ccpar{e^{i  k a \abs{n+m}} - e^{i  k a \abs{n-m}}}, & $n \times m>0$, \\
 \frac{1}{6} \sum_{l=1}^6 \frac{e^{2 \pi i \frac{l (n-m)}{6}} }{E+ \varepsilon_0 + 2 t_{\mathrm{b}} \cos \ccpar{ \frac{2 \pi l}{6} }}, & $n,m=b_{i,j}$, \\
 0, & otherwise,
\end{numcases} 
where $i,j=1,\dots,6$, from where we find
\begin{equation}
 \braketop{n}{G_0}{-1} = -\frac{e^{-i k a n}}{t}
\end{equation} 
and
\begin{equation}
 \braketop{n}{G_0}{1} = -\frac{e^{i  k a n}}{t}.
\end{equation} 
Through these results, we get
\begin{small}
\begin{numcases}{\braket{n}{\psi}= \label{eq:green4:psi05}}
  \sqrt{\frac{2}{N+1}} \sqpar{ \frac{e^{i  k a n}}{2 i} + \frac{e^{- i k a n}}{2 i} \ccpar{-1+ 2 i Y_{\mathrm{L}}^2 \sqrt{1-Z^2} t\braketop{b_1}{G}{b_1} }}, & $n<0$, \\
  \sqrt{\frac{2}{N+1}} \frac{e^{i  k a n}}{2 i} \sqpar{ 2 i Y_{\mathrm{L}} Y_{\mathrm{R}} \sqrt{1-Z^2} t \braketop{b_4}{G}{b_1} }, & $n>0$.
\end{numcases} 
\end{small}

Finally, we only need the elements $\braketop{b_1}{G}{b_1}$ and $\braketop{b_4}{G}{b_1}$. 
These elements can be
calculated through the system
\begin{numcases}{}
 \braketop{b_1}{G}{b_1} = \braketop{b_1}{G_0}{b_1} -\alpha_{\mathrm{L}} \braketop{b_1}{G_0}{b_1}\braketop{-1}{G}{b_1} -\alpha_{\mathrm{R}} \braketop{b_1}{G_0}{b_4}\braketop{1}{G}{b_1}, \\
 \braketop{b_4}{G}{b_1} = \braketop{b_4}{G_0}{b_1} -\alpha_{\mathrm{L}} \braketop{b_4}{G_0}{b_1}\braketop{-1}{G}{b_1} -\alpha_{\mathrm{R}} \braketop{b_4}{G_0}{b_4}\braketop{1}{G}{b_1}, \\
 \braketop{-1}{G}{b_1} = -\alpha_{\mathrm{L}} \braketop{-1}{G_0}{-1}\braketop{b_1}{G}{b_1} \\
 \braketop{1}{G}{b_1} = -\alpha_{\mathrm{R}} \braketop{1}{G_0}{1}\braketop{b_4}{G}{b_1},
\end{numcases} 
\noindent where all equations were derived from the Dyson equation. 
It follows that, defining $G_{n,m} \equiv \braketop{n}{G}{m}$ 
and $G^0_{n,m} \equiv \braketop{n}{G_0}{m}$, we get
\begin{footnotesize}
\begin{equation}
G_{b_1,b_1} = \frac{G^0_{b_1,b_1} + \alpha_{\mathrm{R}}^2  G^0_{1,1} \ccpar{G^0_{b_1,b_4}G^0_{b_4,b_1}-G^0_{b_1,b_1}G^0_{b_4,b_4}}}{1-\alpha_{\mathrm{L}}^2 G^0_{-1,-1}G^0_{b_1,b_1}-\alpha_{\mathrm{R}}^2 G^0_{1,1}G^0_{b_4,b_4}-\alpha_{\mathrm{L}}^2\alpha_{\mathrm{R}}^2 G^0_{-1,-1} G^0_{1,1} \ccpar{G^0_{b_1,b_4}G^0_{b_4,b_1}-G^0_{b_1,b_1}G^0_{b_4,b_4}}}
 \label{eq:green4:Gb1b1}
\end{equation} 
\end{footnotesize}
and
\begin{footnotesize}
\begin{equation}
G_{b_4,b_1} = \frac{G^0_{b_4,b_1}}{1-\alpha_{\mathrm{L}}^2 G^0_{-1,-1}G^0_{b_1,b_1}-\alpha_{\mathrm{R}}^2 G^0_{1,1}G^0_{b_4,b_4}-\alpha_{\mathrm{L}}^2\alpha_{\mathrm{R}}^2 G^0_{-1,-1} G^0_{1,1} \ccpar{G^0_{b_1,b_4}G^0_{b_4,b_1}-G^0_{b_1,b_1}G^0_{b_4,b_4}}}.
 \label{eq:green4:Gb4b1}
\end{equation}
\end{footnotesize}

To close the problem, we only need some matrix elements of $G_0$, which can be found using again the result from equation \eqref{eq:green4:G0nm}.
Doing so, and defining $\Upsilon \equiv {t_{\mathrm{b}}}/{t}$, we get
\begin{equation}
 G^0_{-1,-1}=G^0_{1,1} = - \frac{e^{i  k a}}{t} = \frac{1}{t} \sqpar{Z - i\sqrt{1-Z^2}},
\end{equation}
\begin{equation}
 G^0_{b_1,b_1}=G^0_{b_4,b_4} = \frac{1}{t} \left\{\frac{-\frac{1}{2\Upsilon^3}\ccpar{Z+\frac{X}{2}}\sqpar{\frac{1}{\Upsilon^2}\ccpar{Z+\frac{X}{2}}^2 - 3}}{ \sqpar{\frac{1}{\Upsilon^2}\ccpar{ Z+\frac{X}{2} }^2-1}\sqpar{\frac{4}{\Upsilon^2}\ccpar{ Z+\frac{X}{2} }^2-1}}\right\},
\end{equation}
and
\begin{equation}
 G^0_{b_4,b_1}= G^0_{b_1,b_4} = \frac{1}{t} \left\{\frac{\frac{1}{2\Upsilon}}{ \sqpar{\frac{1}{\Upsilon^2}\ccpar{ Z+\frac{X}{2} }^2-1}\sqpar{\frac{4}{\Upsilon^2}\ccpar{ Z+\frac{X}{2} }^2-1}}\right\}.
\end{equation}

Note that all these elements are only numbers (real or complex), that are function of the energy and
several other parameters, but not the position $n$. As a consequence, the same is true for the elements
$G_{b_1,b_1}$ and $G_{b_4,b_1}$. This means that we find
\begin{numcases}{\label{eq:green4:coefs}}
  r= 2 i Y_{\mathrm{L}}^2 \sqrt{1-Z^2} \sqpar{ t G_{b_1,b_1} } -1, \\
  \tau= 2 i Y_{\mathrm{L}} Y_{\mathrm{R}} \sqrt{1-Z^2} \sqpar{ t G_{b_4,b_1} },
\end{numcases}

\noindent and, finally,
\begin{numcases}{\label{eq:green4:coefs2}}
  \mathcal{R}= 1 + 4 Y_{\mathrm{L}}^4 \ccpar{1-Z^2} t^2 \abs{ G_{b_1,b_1} }^2 + 4 Y_{\mathrm{L}}^2 \sqrt{1-Z^2} \mathrm{Im} \sqpar{ t G_{b_1,b_1} },  \\
  \mathcal{T}= 4 \ccpar{Y_{\mathrm{L}} Y_{\mathrm{R}}}^2 \ccpar{1-Z^2} t^2 \abs{ G_{b_4,b_1} }^2. 
\end{numcases} 

\begin{figure}[htbp]
\centering
\includegraphics[width=\textwidth]{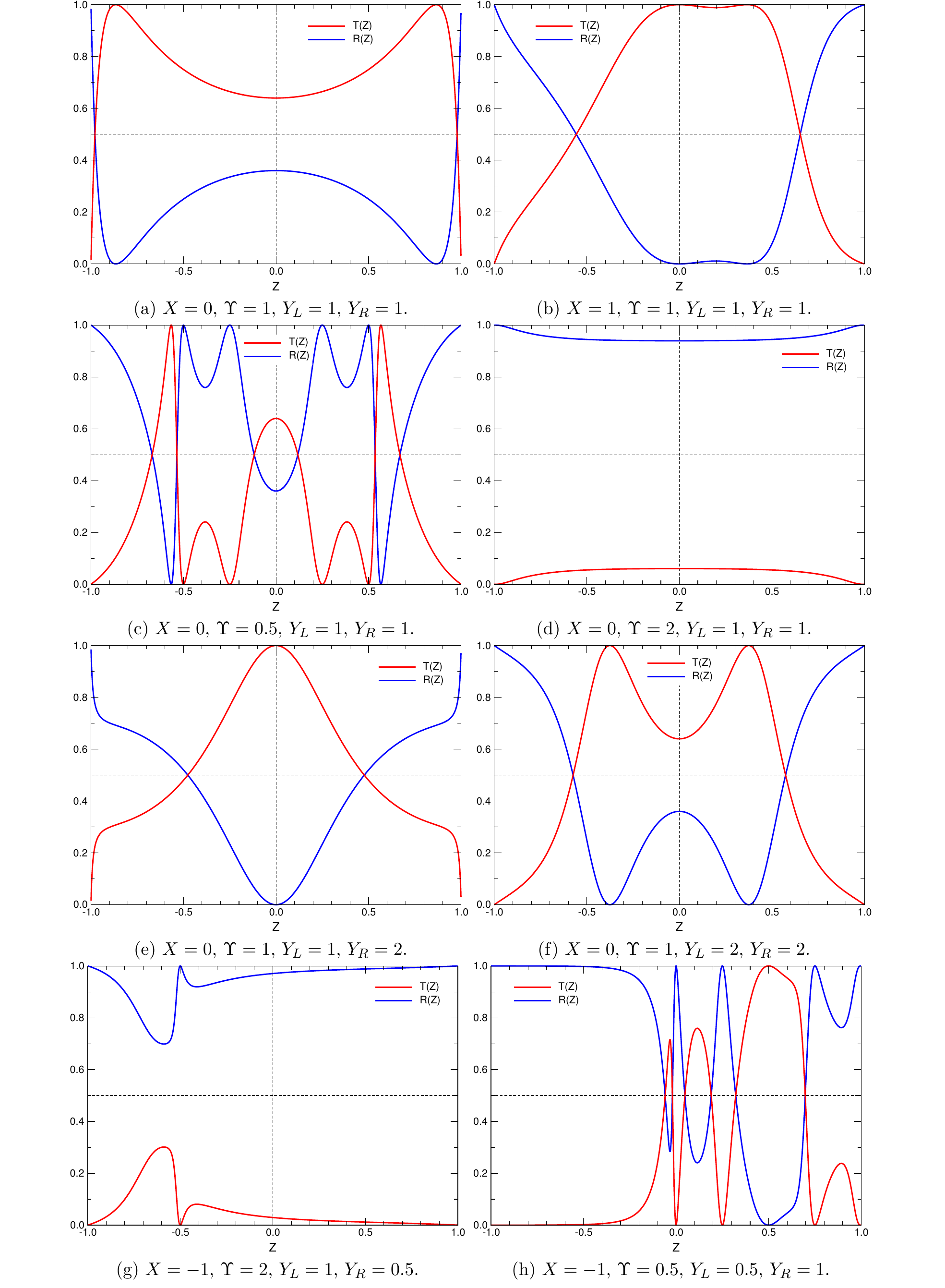}
\caption{Graphic representations of the transmittance and reflectance through a benzene ring, as a 
function of energy, for different values of the parameters $X$, $\Upsilon$, $Y_{\mathrm{L}}$ and $Y_{\mathrm{R}}$.}
\label{fig:TR_benzene}
\end{figure}

In figure \ref{fig:TR_benzene}, we present some graphical representations of $\mathcal{R}$ and $\mathcal{T}$
as function of the energy $Z$, for several combinations of the parameters
$X$, $\Upsilon$, $Y_{\mathrm{L}}$ and $Y_{\mathrm{R}}$. The discussion of the results is left for the 
final section of the paper.

\subsection{Local Density of States}
\label{sec:LDOS}

One of the main advantages of the Green's function approach is that the calculation of the Local Density
of States (LDOS) in the atom of the position $n$ is straightforward. It is given by \cite{Economou}
\begin{equation}
 \rho_n(Z) = - \frac{1}{\pi} \mathrm{Im} G_{n,n}(Z),
\end{equation} 
\noindent where $G_{n,n}\equiv \braketop{n}{G}{n}$. For the benzene system, we are interested in the LDOS
in each of the atoms of the benzene ring. In general, it should be different for each atom except, due to
the symmetry of the system, $\rho_{b_2}(Z)=\rho_{b_6}(Z)$ and $\rho_{b_3}(Z)=\rho_{b_5}(Z)$. In the particular
case when $\alpha_{\mathrm{L}}=\alpha_{\mathrm{R}}$, there is a higher level of symmetry which imposes that $\rho_{b_2}(Z)=\rho_{b_3}(Z)=\rho_{b_5}(Z)=\rho_{b_6}(Z)$
and also $\rho_{b_1}(Z)=\rho_{b_4}(Z)$. 


To calculate each of this different local densities of states, we need to find the matrix elements $\braketop{b_n}{G}{b_n}$, 
with $n=1,...,6$. Through Dyson equation, we find
\begin{numcases}{}
 \braketop{b_n}{G}{b_n} = \braketop{b_n}{G_0}{b_n} - \alpha_{\mathrm{L}} \braketop{b_n}{G_0}{b_1}\braketop{-1}{G}{b_n} - \alpha_{\mathrm{R}} \braketop{b_n}{G_0}{b_4}\braketop{1}{G}{b_n}, \\
 \braketop{ -1}{G}{b_n} = - \alpha_{\mathrm{L}} \braketop{-1}{G_0}{-1}\braketop{b_1}{G}{b_n}, \\
 \braketop{  1}{G}{b_n} = - \alpha_{\mathrm{R}} \braketop{ 1}{G_0}{ 1}\braketop{b_4}{G}{b_n}, \\
 \braketop{b_1}{G}{b_n} = \braketop{b_1}{G_0}{b_n} - \alpha_{\mathrm{L}} \braketop{b_1}{G_0}{b_1}\braketop{-1}{G}{b_n} - \alpha_{\mathrm{R}} \braketop{b_1}{G_0}{b_4}\braketop{1}{G}{b_n}, \\
 \braketop{b_4}{G}{b_n} = \braketop{b_4}{G_0}{b_n} - \alpha_{\mathrm{L}} \braketop{b_4}{G_0}{b_1}\braketop{-1}{G}{b_n} - \alpha_{\mathrm{R}} \braketop{b_4}{G_0}{b_4}\braketop{1}{G}{b_n}.
\end{numcases}

\begin{figure}[htbp]
\centering
\includegraphics[width=\textwidth]{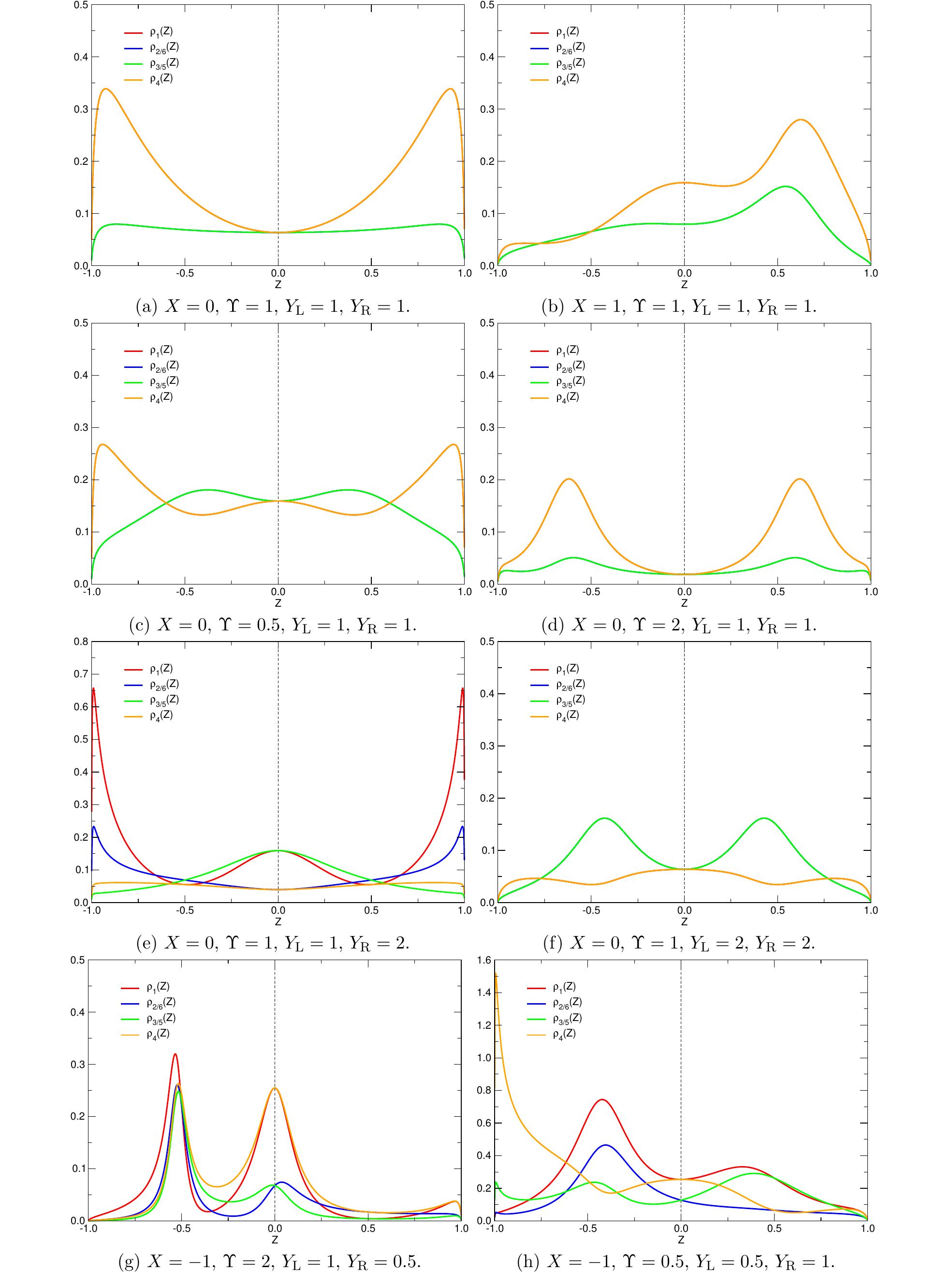}
\caption{Graphic representations of the local density of states in the atoms of a benzene ring, as a 
function of energy, for different values of the parameters $X$, $\Upsilon$, $Y_{\mathrm{L}}$ and $Y_{\mathrm{R}}$. Note that,
in figures (a)--(d) and (f), the red and orange lines are superimposed, as well as the blue and green 
ones.}
\label{fig:LDOS_benzene}
\end{figure}

\noindent These five equations make up a closed system from where it is possible to find the value of $\braketop{b_n}{G}{b_n}$. 
However, the analytic solutions, although possible to find, are too complex to have any interest in being presented. As an
alternative, we present, in figure \ref{fig:LDOS_benzene}, graphical representations of the LDOS in each atom, as function 
of the energy $Z$, for several combinations of the parameters 
$X$, $\Upsilon$, $Y_{\mathrm{L}}$ and $Y_{\mathrm{R}}$. From the study of this figure it is clear that
the behaviour of the LDOS is rather complex, depending on the actual values of the parameters.

\section{Discussion}

In conclusion, we have developed a formalism, based on Green's functions, for describing the 
electronic transport through molecules represented by tight-binding models. We started by
explaining the general method, defining some useful operators. To illustrate its usage, we
then applied it to the case of the resonant level system, where we were able to find 
its transmittance. To do so, we also found the wave functions and the allowed energies of electrons
propagating in an ideal unidimensional chain. From the expressions found for the transmittance
and reflectance of this system, and the subsequent presented plots, we draw a number of useful results:
\begin{itemize}
\item both $\mathcal{T}$ and $\mathcal{R}$ are symmetrical with respect to the line 
$\mathcal{T} = \mathcal{R} = 0.5$. This is due to the condition $\mathcal{T}+\mathcal{R}=1$;

\item in most cases, at $Z=\pm 1$, we found that $\mathcal{T}=0$ and $\mathcal{R}=1$. The exception 
was the plot presented in figure \ref{fig:TR_resonant_level}(h), in which $\mathcal{T}=1$ and 
$\mathcal{R}=0$ at $Z=1$. Through an analysis of the limits
of \eqref{eq:green3:coefs2}, we found that this exceptions occur whenever one set of parameters 
respects the condition $\abs{X/2}=\abs{(Y_{\mathrm{L}}^2+Y_{\mathrm{R}}^2)/2-1}$ [which is the case of the parameters in figure 
\ref{fig:TR_resonant_level}(h)]. In these cases, the transmittance is total in at least one
of the extremes of the band;

\item when $X=0$, the these functions are also symmetrical with respect to $Z=0$. This means
that they must have a maximum or minimum at this energy. On the other hand, when $X$ was a
positive (negative) number, the peak tended to slide to the right (left), for the same parameters
$Y_{\mathrm{L}}$ and $Y_{\mathrm{R}}$. Analysing \eqref{eq:green3:coefs2}, we found that the extreme of these functions
depends on the other parameters through the expression
\begin{numcases}{Z_{\mathrm{extremo}}=}
\frac{\frac{X}{2}}{\frac{Y_{\mathrm{L}}^2+Y_{\mathrm{R}}^2}{2}-1}, & $\abs{\frac{X}{2}}\leq\abs{\frac{Y_{\mathrm{L}}^2+Y_{\mathrm{R}}^2}{2}-1}$, \\
\frac{\frac{Y_{\mathrm{L}}^2+Y_{\mathrm{R}}^2}{2}-1}{\frac{X}{2}}, & $\abs{\frac{X}{2}} >  \abs{\frac{Y_{\mathrm{L}}^2+Y_{\mathrm{R}}^2}{2}-1}$.
\end{numcases}
This result explains both the behaviour of the peak with a change in $X$ and also the exception presented
in the previous point;

\item through the two previous points, we conclude that the extreme found must be a maximum of the
transmittance and a minimum of the reflectance. Using \eqref{eq:green3:coefs2}, we obtain that
\cite{footnote2}
\begin{numcases}{\mathcal{T}_{\mathrm{max}}= \label{eq:maxtran1}} 
\frac{4Y_{\mathrm{L}}^2Y_{\mathrm{R}}^2}{\ccpar{Y_{\mathrm{L}}^2+Y_{\mathrm{R}}^2}^2}, & $\abs{\frac{X}{2}}\leq\abs{\frac{Y_{\mathrm{L}}^2+Y_{\mathrm{R}}^2}{2}-1}$, \\
\frac{Y_{\mathrm{L}}^2Y_{\mathrm{R}}^2}{\ccpar{\frac{X}{2}}^2 + Y_{\mathrm{L}}^2+Y_{\mathrm{R}}^2 -1}, & $\abs{\frac{X}{2}} >  \abs{\frac{Y_{\mathrm{L}}^2+Y_{\mathrm{R}}^2}{2}-1}$.
\end{numcases}
In particular, when $Y_{\mathrm{L}}=Y_{\mathrm{R}} \equiv Y$, we get the important result
\begin{numcases}{\mathcal{T}_{\mathrm{max}}= \label{eq:maxtran2}} 
1, & $\abs{\frac{X}{2}} \leq \abs{Y^2-1}$, \\
\frac{Y^4}{\ccpar{\frac{X}{2}}^2 + 2Y^2 -1}, & $\abs{\frac{X}{2}} > \abs{Y^2-1}$.
\end{numcases}
This result is important because, as we can see, when $Y_{\mathrm{L}}=Y_{\mathrm{R}}$, we can adjust the parameters $X$ and $Y$
so that, at some energy, we have maximum transmittance. This is not possible when $Y_{\mathrm{L}}\neq Y_{\mathrm{R}}$, nor in
the simplest case of point defect in which $Y_{\mathrm{L}}=Y_{\mathrm{R}}=1$. 
This is why this problem is called the level resonant
system. In the diagram of figure \ref{fig:phasespace}, the region of the phase space $XY$ where
$\mathcal{T}_{\mathrm{max}}=1$ is presented in grey. On the other hand, in the white region of the same diagram,
the maximum transmittance is always lower than one, decreasing with the distance to the bold line
defined by $\abs{X/2} = \abs{Y^2-1}$. To understand the topology of this region, 
some level lines are represented;

\begin{figure}[htbp]
\centering
\includegraphics{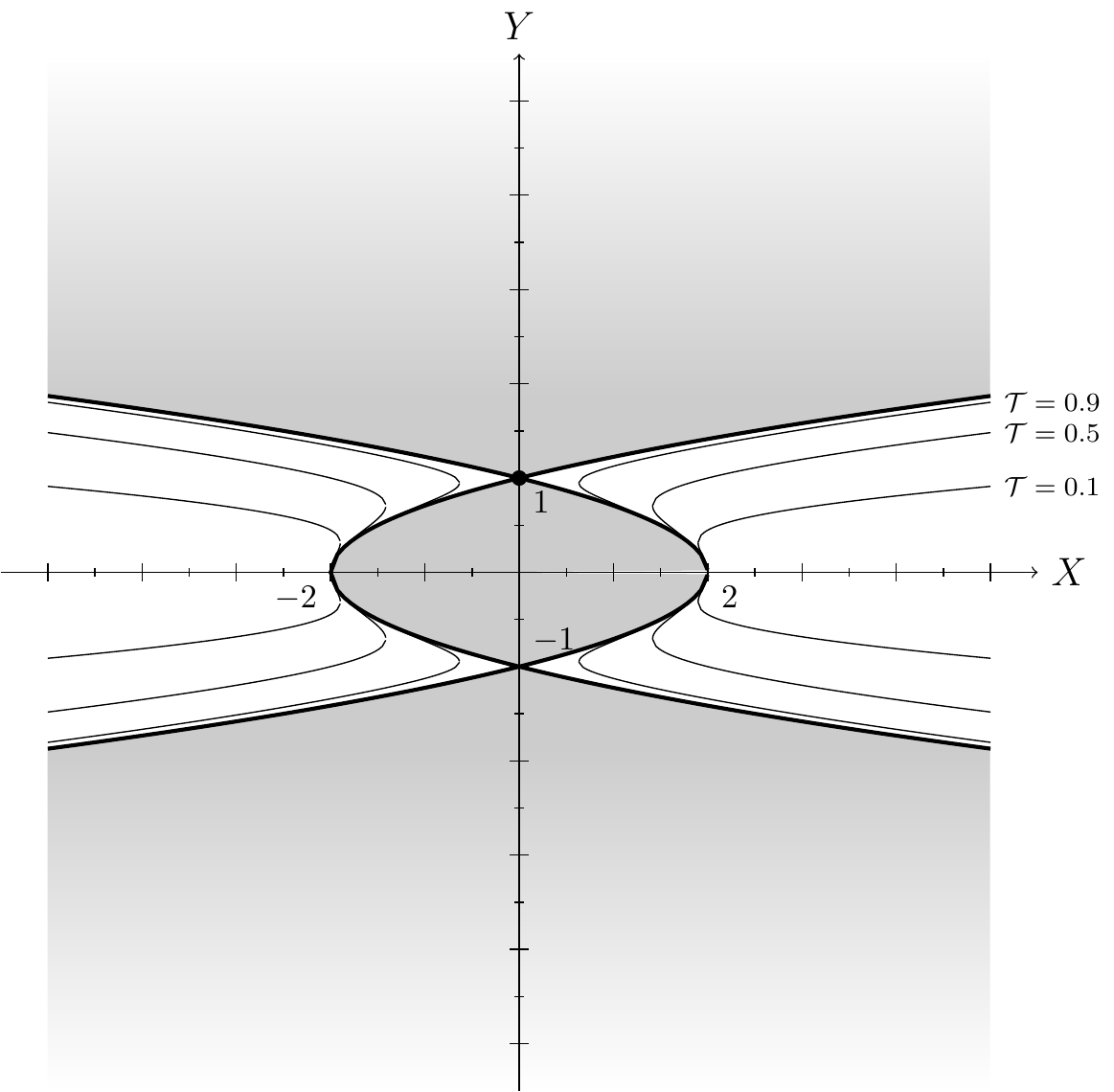}
\caption{This diagram shows: in a bold line, the points for which $\abs{X/2} = \abs{Y^2-1}$; 
in a grey shadow, the region of the phase space $XY$ where $\mathcal{T}_{\mathrm{max}}=1$; in white,
the region where $\mathcal{T}_{\mathrm{max}}<1$; in thin lines, some level lines, for $\mathcal{T}_{\mathrm{max}}=0.1$,
$\mathcal{T}_{\mathrm{max}}=0.5$ and $\mathcal{T}_{\mathrm{max}}=0.9$; and a black dot which represents
the point in the phase space corresponding to the ideal system, that is, $Y=1$ and $X=0$. The grey shaded 
region continues as $Y \rightarrow \pm \infty$. Note that
the diagram is symmetrical both with respect to the lines $X=0$ and $Y=0$.}
\label{fig:phasespace}
\end{figure}

\item in the general case, for the same $Y_{\mathrm{L}}$ and $Y_{\mathrm{R}}$, the transmittance increases with the decreasing of $X$, 
as expected, but only when $\abs{X/2} >  \abs{(Y_{\mathrm{L}}^2+Y_{\mathrm{R}}^2)/2-1}$. Elsewhere, this value is independent of $X$. 
On the other hand, for the same $X$, the transmittance tends to increase when $Y_{\mathrm{L}}=Y_{\mathrm{R}}$ and also when 
$Y_{\mathrm{L}}$ and $Y_{\mathrm{R}}$, as represented in the phase space $XY_{\mathrm{L}}Y_{\mathrm{R}}$, get closer 
to the surface defined by $\abs{X/2} = \abs{(Y_{\mathrm{L}}^2+Y_{\mathrm{R}}^2)/2-1}$, and decrease with the distance to 
that surface. In particular, it is null when either $Y_{\mathrm{L}}$ or $Y_{\mathrm{R}}$ are null.

\end{itemize}

We then studied the case of the benzene system, 
and presented  plots of its transmittance and reflectance, for different sets of 
parameters. Here, the plots obtained for the transmittance 
were much more complex, depending on four parameters,
which means that a study like the one presented for the resonant level system above is much more
difficult to elaborate. Indeed, It would require an extensive study of the limits of the transmittance and
reflectance for several values of the presented parameters, and also the production of much 
more plots than the ones presented here, what
would worth of paper of itself. Our goal was rather to show that the Green's function method an
easily be applied to a problem of moderate difficulty.
The same can be said about the local density 
of states in each atom of the benzene ring, which was also calculated in this paper.

Nevertheless, some qualitative aspects can be drawn from the results of Fig. \ref{fig:TR_benzene}.
Taking the single example of panel (h) of that figure we seen that the transmittance is strongly 
suppressed for values of the energy such that $Z<-0.5$. If we now imagine that the energy of 
the electrons can be tuned, such a system would correspond to an electronic filter, where
only the electrons with energies in the range $-0.5<Z<0$ would be transmitted. The same can be said
for the results in panel (g) and energies $Z>0$.

This method, as should now be clear, can be easily generalized to more complex problems, such 
as a chain of atoms, like a polyacetylene, or two (or more) benzene rings in a row. However, we 
can conclude that the analytical expressions of the found quantities, although possible to find
through this method, can get rather complex when the complexity of the problem increases,
due to the increase of the number of atoms involved in the calculation. In this case, the method
can be formulated using efficient numerical methods. 

In synthesis, as should now be obvious, the Green's functions 
method is ideal for tackling transport problems in molecular electronics, providing, in many cases,
analytical expressions for the quantum transmittance, which can further 
be analysed in detail for the benefit of gaining insight  on the transport properties 
of a given system.

\newpage
\setcounter{section}{1}
\appendix
\section{Matrix elements of the free Green's operator for the resonant level system}
\label{sec:anexoG03}

In this appendix, we present the calculation of the matrix elements $G_0(n,m)$ of the free Green's
function operator for the resonant level system. These elements are 
given by
\begin{eqnarray}
 G_0(n,m) = \braketop{n}{G_0}{m} &= \bra{n} \frac{1}{E-H_0+i\epsilon} \ket{m} 
 \label{eq:G03:green1}
\end{eqnarray} 

\noindent where $H_0 = H_{\mathrm{L}} + H_{\mathrm{C}} + H_{\mathrm{R}}$, defined as in \eqref{eq:green3:HL}--\eqref{eq:green3:HR}. We start by
using a Taylor expansion of the form $1/(x_0-x)=\sum_{l=0}^{+\infty} x^l/x_0^{l+1}$, so that
\begin{eqnarray}
 G_0(n,m) &= \sum_{l=0}^{+\infty} \bra{n} \frac{\ccpar{H_{\mathrm{L}}+H_{\mathrm{C}}+H_{\mathrm{R}}}^l}{\ccpar{E+i\epsilon}^{l+1}} \ket{m}.
 \label{eq:G03:green2}
\end{eqnarray}

\noindent The subsequent expansion of the power in the numerator, for any $l$, will mix the different components
of $H_0$ in many possible combinations. However, since the three components $H_{\mathrm{L}}$, $H_{\mathrm{C}}$ and $H_{\mathrm{R}}$ are
uncoupled, any operator operating in a position state of a different region will vanish (and, once these
particular operators are Hermitian, this is valid for both the bra and the ket surrounding it), remaining only
\begin{footnotesize}
\begin{numcases}{G_0(n,m) =}
 \sum_{l=0}^{+\infty} \frac{\braketop{n}{H_{\mathrm{L}}^l}{m}}{\ccpar{E+i\epsilon}^{l+1}} = \bra{n} \frac{1}{E-H_{\mathrm{L}}+i\epsilon} \ket{m} \equiv \bra{n} G_{\mathrm{L}} \ket{m} , & $n,m<0$, \\
 \sum_{l=0}^{+\infty} \frac{\braketop{n}{H_{\mathrm{C}}^l}{m}}{\ccpar{E+i\epsilon}^{l+1}} = \bra{n} \frac{1}{E-H_{\mathrm{C}}+i\epsilon} \ket{m} \equiv \bra{n} G_{\mathrm{C}} \ket{m} , & $n,m=0$, \\
 \sum_{l=0}^{+\infty} \frac{\braketop{n}{H_{\mathrm{R}}^l}{m}}{\ccpar{E+i\epsilon}^{l+1}} = \bra{n} \frac{1}{E-H_{\mathrm{R}}+i\epsilon} \ket{m} \equiv \bra{n} G_{\mathrm{R}} \ket{m} , & $n,m>0$, \\
 0, & otherwise.
\end{numcases}
\end{footnotesize}

\noindent As we can see now, the value of the matrix elements will, in general, be different for three different regions,
so we shall now proceed to each calculation separately.

\subsection{Calculation for the 1D chains next to the defect}
\label{sec:G03:LR}

Due to the high symmetry verified between the chains in both sides of the defect, these two cases will
be treated simultaneously. In these regions, we ought to find the elements
\begin{numcases}{G_0(n,m)=}
 G_{\mathrm{L}}(n,m) = \braketop{n}{G_{\mathrm{L}}}{m}, & $n,m<0$, \\
 G_{\mathrm{R}}(n,m) = \braketop{n}{G_{\mathrm{R}}}{m}, & $n,m>0$.
\end{numcases} 
\noindent where
\begin{numcases}{}
 G_{\mathrm{L}} = \frac{1}{E-H_{\mathrm{L}}+i\epsilon}, \\
 G_{\mathrm{R}} = \frac{1}{E-H_{\mathrm{R}}+i\epsilon}.
\end{numcases} 

Back in section 2, the eigenstates of $H_{\mathrm{L}}$ and $H_{\mathrm{R}}$ were found, and are presented in \eqref{eq:green3:psiL} and
\eqref{eq:green3:psiR}. Since each of them makes up a complete basis, we can use the notation $\ket{\psi_{\mathrm{L}}} \equiv \ket{k_{\mathrm{L}}}$
and $\ket{\psi_{\mathrm{R}}} \equiv \ket{k_{\mathrm{R}}}$ to write
\begin{eqnarray}
 \mathbb{I} &= \sum_k \ket{k_{\mathrm{L}}}\bra{k_{\mathrm{L}}}
            = \sum_k \ket{k_{\mathrm{R}}}\bra{k_{\mathrm{R}}}
\end{eqnarray} 

\noindent where $\mathbb{I}$ is the unity operator and the sums are over all allowed values of $k$. With this result, 
the matrix elements can be written as
\begin{numcases}{G_0(n,m) =}
 \sum_k \braket{n}{k_{\mathrm{L}}}\braketop{k_{\mathrm{L}}}{G_{\mathrm{L}}}{m} = \sum_k \sum_{l=0}^{+\infty} \frac{\braketop{k_{\mathrm{L}}}{H_{\mathrm{L}}^l}{m}}{\ccpar{E+i\epsilon}^{l+1}} \braket{n}{k_{\mathrm{L}}}, & $n,m<0$, \\
 \sum_k \braket{n}{k_{\mathrm{R}}}\braketop{k_{\mathrm{R}}}{G_{\mathrm{R}}}{m} = \sum_k \sum_{l=0}^{+\infty} \frac{\braketop{k_{\mathrm{R}}}{H_{\mathrm{R}}^l}{m}}{\ccpar{E+i\epsilon}^{l+1}} \braket{n}{k_{\mathrm{R}}}, & $n,m>0$.
\end{numcases} 

The advantage of writing these elements as such is that $H_{\mathrm{L}}^l \ket{k_{\mathrm{L}}} = E(k)^l \ket{k_{\mathrm{L}}}$, 
with $E(k)$ defined in \eqref{eq:green3:EL}. The same is true for $\ket{k_{\mathrm{R}}}$ and $H_{\mathrm{R}}$. Through this result,
the previous expression simplifies to
\begin{numcases}{G_0(n,m) =}
 \sum_k \frac{\braket{n}{k_{\mathrm{L}}}\braket{k_{\mathrm{L}}}{m}}{E + 2 t \cos \ccpar{ k a }+i\epsilon}, & $n,m<0$, \\
 \sum_k \frac{\braket{n}{k_{\mathrm{R}}}\braket{k_{\mathrm{R}}}{m}}{E + 2 t \cos \ccpar{ k a }+i\epsilon}, & $n,m>0$.
\end{numcases} 

Now, if we recall \eqref{eq:k1D}, we see that $\Delta k \equiv k_{n+1} - k_n = \pi\sqpar{a \ccpar{N+1}}^{-1}$. When $N \gg 1$, then
$\Delta k$ tends to an infinitesimal element $dk$, and thus, if we multiply (and divide) the previous equations by $\Delta k$, we
can approximate the sum to an integral. Moreover, we already know the value of the operations $\braket{n}{k}$ and $\braket{k}{n}$ 
($k=k_{\mathrm{L}},k_{\mathrm{R}}$). Through these two results, we find that, for both sides of the defect, the matrix
elements of $G_0$ are given by
\begin{eqnarray}
 G_0(n,m) &= \frac{1}{\Delta k} \frac{2}{N+1} \int_{0}^{\frac{\pi}{a}} dk \frac{\sin \ccpar{k a n} \sin \ccpar{k a m}}{E + 2t \cos \ccpar{ka}+i\epsilon} \\
          &= \frac{2a}{\pi} \frac{1}{2} \int_{-\frac{\pi}{a}}^{\frac{\pi}{a}} dk \frac{\sin \ccpar{k a n} \sin \ccpar{k a m}}{E + 2t \cos \ccpar{ka}+i\epsilon},
\end{eqnarray} 

\noindent where we noted that the function being integrated is even. This expression is now valid when $n,m>0$ or $n,m<0$, or, to
simplify, when $n \times m >0$. Writing now the sine functions in the Euler notation, we get
\begin{equation}
G_0(n,m) = -\frac{a}{4\pi} \int_{-\frac{\pi}{a}}^{\frac{\pi}{a}} dk \frac{ e^{i k a (n+m)} - e^{i k a (n-m)} + e^{-i k a (n+m)} - e^{-i k a (n-m)}   }{E + 2t \cos \ccpar{ka}+i\epsilon}.
\end{equation} 

\noindent However, due to the parity of the functions sine and cosine that make up each complex exponential in the previous equation, we verify
that the integration of the terms whose numerator is $e^{i k a (n+m)} - e^{i k a (n-m)}$ is equal to ones whose
numerator is $e^{-i k a (n+m)} - e^{-i k a (n-m)}$, which means the integral can be rewritten in the simpler form
\begin{eqnarray}
 G_0(n,m) &= -\frac{a}{4\pi} 2 \int_{-\frac{\pi}{a}}^{\frac{\pi}{a}} dk \frac{e^{i k a \abs{n+m}} - e^{i k a \abs{n-m}}}{E + 2t \cos \ccpar{ka}+i\epsilon} \\
          &= \frac{a}{2\pi} \sqpar{ \int_{-\frac{\pi}{a}}^{\frac{\pi}{a}} dk \frac{e^{i k a \abs{n-m}}}{E + 2t \cos \ccpar{ka}+i\epsilon} - \int_{-\frac{\pi}{a}}^{\frac{\pi}{a}} dk \frac{e^{i k a \abs{n+m}}}{E + 2t \cos \ccpar{ka}+i\epsilon} }\\
          & \equiv  I(\abs{n-m}) - I(\abs{n+m}),
\end{eqnarray} 

\noindent where
\begin{equation}
 I(x) \equiv  \frac{a}{2\pi} \int_{-\frac{\pi}{a}}^{\frac{\pi}{a}} dk \frac{e^{i k a x}}{E + 2t \cos \ccpar{ka}+i\epsilon}.
\end{equation}

\noindent Thus, we need now to find the general solution of $I(x)$ to find the matrix elements we need. The
solution of this integral was found to be (see below)
\begin{equation}
 I(x) =  \frac{-i e^{i k a x}}{2t\sqrt{1-\ccpar{\frac{E}{2t}}^2}},
\end{equation} 

\noindent what means that the general result for the matrix elements of $G_0$, for $n\times m>0$, is
\begin{equation}
 G_0(n,m) = \frac{i}{2t\sqrt{1-\ccpar{\frac{E}{2t}}^2}} \ccpar{e^{i k a \abs{n+m}} - e^{i k a \abs{n-m}}}.
 \label{eq:G0:chains}
\end{equation}

We need yet to prove the solution of the integral $I(x)$. Defining $\theta \equiv ka$ and $z=E+i \epsilon$,
we rewrite the integral as
\begin{equation}
 I(x) = \frac{1}{2 \pi} \int_{-\pi}^{\pi} d\theta \frac{e^{i \theta x}}{z+t\ccpar{e^{i \theta}+e^{-i \theta}}}.
\end{equation} 
Defining also  $w\equiv e^{i x}$, the previous integral takes the form
\begin{eqnarray}
  I(x) &= \frac{1}{2 \pi} \oint_{\gamma} \frac{dw}{iw} \frac{w^{x}}{z+t(w+w^{-1})} \\
           &= \frac{1}{2 \pi i t} \oint_{\gamma} dw \frac{w^{x}}{ w^2 + \frac{w z}{t} + 1 } \\ \label{eq:anexoD:integral1}
           &= \frac{1}{2 \pi i t} \oint_{\gamma} dw \frac{w^{x}}{ (w-w_1)(w-w_2) },
\end{eqnarray} 

\noindent where $\gamma$ is an unitary circle in the complex plane and $w_1$ e $w_2$ are the two roots
of $w^2 + w z/t + 1$, given by
\begin{equation}
 w_{1,2} = -\frac{z}{2t} \pm i\sqrt{1-\ccpar{\frac{z}{2t}}^2}
 \label{eq:polos}
\end{equation} 

\noindent (where the subscript $1$ corresponds to the signal $+$, and vice versa). Note that we
are interested in the case when $-2t < E < 2t$ (what corresponds to the bandwidth of these chains). 
The integral in now written in such a way that it is now convenient to use the Residue Theorem. However,
to do so, we need to know whether each singularity is contained in $\gamma$. That calculation is done in \ref{sec:anexo:modulo},
from where it was concluded that, when $\epsilon \to 0^+$, then ${w_1}\to 1^-$ and ${w_2}\to 1^+$. This means that
\begin{eqnarray}
 I(x) &= \frac{1}{2 \pi i t} 2 \pi i \mathrm{Res}_{w=w_1}\sqpar{\frac{w^{x}}{ (w-w_1)(w-w_2) }} \\
          &= \frac{1}{t} \frac{w_1^{x}}{ w_1-w_2 }, \label{eq:anexoD:integral2}
\end{eqnarray} 
\noindent where $\mathrm{Res}_{z=z_0}f(z)$ denotes the residue of $f(z)$ at $z=z_0$. Note now that, when $\epsilon \to 0$,
we can use the relation $E=-2t 	\cos \ccpar{ka}$ to write the singularities as
\begin{equation}
 w_{1,2} = -\frac{E}{2t} \pm i \sqrt{1 - \ccpar{\frac{E}{2t}}^2} = e^{\pm i k a}.  \label{eq:anexoD:w12dentro}
\end{equation}

\noindent Through this result, we finally obtain the already stated result
\begin{equation}
 I(x) = \frac{-i e^{i k a x}}{2t\sqrt{1-\ccpar{\frac{E}{2t}}^2}}.
 \label{eq:G0:resI(x)}
\end{equation} 

\subsection{Calculation for the defect}
\label{sec:G03:C}

In this region, we need to find the elements
\begin{equation}
G_{\mathrm{C}}(n,m) = \braketop{n}{G_{\mathrm{C}}}{m},
\end{equation} 
\noindent where
\begin{equation}
 G_{\mathrm{C}} = \frac{1}{E-H_{\mathrm{C}}+i\epsilon},
\end{equation} 
and the Hamiltonian $H_{\mathrm{C}}$ was defined in \eqref{eq:green3:HC}. As was stated before, the only matrix 
element of $G_{\mathrm{C}}$ that does not vanish is $G_{\mathrm{C}}(0,0)$. This case is much simpler
than the previous one, because the position state $\ket{0}$ is already an eigenstate of $H_{\mathrm{C}}$,
with $H_{\mathrm{C}}\ket{0} = -\varepsilon_0 \ket{0}$. As such,
\begin{eqnarray}
 G_0(0,0) &=  \sum_{l=0}^{+\infty} \frac{\braketop{0}{H_{\mathrm{C}}^l}{0}}{\ccpar{E+i\epsilon}^{l+1}} \\
          &=  \sum_{l=0}^{+\infty} \frac{\braketop{0}{\ccpar{-\varepsilon_0}^l}{0}}{\ccpar{E+i\epsilon}^{l+1}} \\
          &\rightarrow \frac{1}{E+\varepsilon_0},  \label{eq:anexogreen3:G0(0,0)}
\end{eqnarray} 

\noindent when $\epsilon \to 0$.

In short, the matrix elements of $G_0$ for the resonant level system are
\begin{numcases}{ G_0(n,m) = \label{eq:anexogreen3:G0final}}
 \frac{1}{E+\varepsilon_0}, & $n=m=0$, \\
 \frac{i}{2t\sqrt{1-\ccpar{\frac{E}{2t}}^2}} \ccpar{e^{i k a \abs{n+m}} - e^{i k a \abs{n-m}}}, & $n \times m>0$, \\
 0, & otherwise.
\end{numcases} 

\subsection{Evaluation of the absolute value of the singularities found}
\label{sec:anexo:modulo}

In this appendix, we will evaluate the absolute value of the complex numbers $w_{1,2}$, defined in \eqref{eq:polos}.
Let us consider that $\epsilon$ is so small that
\begin{equation}
 \epsilon^2=0,
\end{equation} 
\begin{equation}
 x \epsilon = \sgn{x} \epsilon
\end{equation} 
\noindent and, as a particular case of the previous,
\begin{equation}
 \epsilon + \epsilon = \epsilon,
\end{equation} 
\noindent where $x$ is some real number and the signal function is defined as
\begin{numcases}{\sgn{x}=}
 +1, & $x>0$, \\
 0, & $x=0$, \\
 -1, & $x<0$.
\end{numcases} 

Expanding now the squared binomial expression present in \eqref{eq:polos}, and noting the previous considerations,
we get
\begin{eqnarray}
 w_{1,2} &= -\ccpar{\frac{E+i \epsilon}{2t}} \pm \sqrt{\ccpar{\frac{E}{2t}}^2-1+i \sgn{E} \epsilon} \\
         &= -\ccpar{\frac{E+i \epsilon}{2t}} \pm \sqrt{\sqpar{1-\ccpar{\frac{E}{2t}}^2}\sqpar{-1+i \sgn{E} \epsilon}} \\
         &= -\ccpar{\frac{E+i \epsilon}{2t}} \pm i \sqrt{1-\ccpar{\frac{E}{2t}}^2} \sqrt{1-i \sgn{E} \epsilon}.
	 \label{eq:anexo:modulo:polos2}
\end{eqnarray}

\noindent Since $\epsilon$ is as small as necessary, we can use a first order Taylor approximation of
the function $\sqrt{1-x}$ at $x \approx 0$, with the result $\sqrt{1+x} \approx 1+{x}/{2}$. Doing so,
the previous expression simplifies to
\begin{eqnarray}
  w_{1,2} &= -\ccpar{\frac{E+i \epsilon}{2t}} \pm i \sqrt{1-\ccpar{\frac{E}{2t}}^2} \sqpar{1-i \sgn{E} \epsilon} \\
          &= -\sqpar{\frac{E}{2t} \mp \sgn{E} \epsilon} - i \sqpar{ \mp \sqrt{1-\ccpar{\frac{E}{2t}}^2} + \epsilon }.
\end{eqnarray}

\noindent Consider now the squared absolute value $w_{1,2}$, given by $\abs{w_{1,2}}^2 = w_{1,2} w_{1,2}^*$. Through
this expression,
\begin{eqnarray}
 \abs{w_{1,2}}^2 &= \sqpar{\frac{E}{2t} \mp \sgn{E} \epsilon}^2 + \sqpar{ \mp \sqrt{1-\ccpar{\frac{E}{2t}}^2} + \epsilon }^2 \\
                 &= \sqpar{ \ccpar{\frac{E}{2t}}^2 \mp \sgn{E}^2 \epsilon } + \sqpar{ 1-\ccpar{\frac{E}{2t}}^2 \mp \epsilon \sqrt{1-\ccpar{\frac{E}{2t}}^2}} \\
                 &= 1 \mp \epsilon.
\end{eqnarray}

\noindent Using again the previously presented Taylor approximation, we get
\begin{equation}
 \abs{w_{1,2}} = 1 \mp \epsilon,
\end{equation} 

\noindent or, writing it in a different way,
\begin{numcases}{}
 \lim_{\epsilon \to 0^+} \abs{w_{1}} = 1^-, \\
 \lim_{\epsilon \to 0^+} \abs{w_{2}} = 1^+.
\end{numcases}

We conclude, therefore, that $w_1$ lies within the unitary circle, unlike $w_2$.

\newpage
\section{Matrix elements of the free Green's operator for the benzene system}
\label{sec:anexoG04}

In this appendix, we present the calculation of the matrix elements $G_0(n,m)$ of the free
Green's operator for the benzene system. The method will be similar
to the one carried out for the case of the resonant level system (\ref{sec:anexoG03}). These
elements are given by
\begin{eqnarray}
 G_0(n,m) \equiv \braketop{n}{G_0}{m} &= \bra{n} \frac{1}{E-H_0+i\epsilon} \ket{m} 
 \label{eq:G04:green1}
\end{eqnarray} 
\noindent where $H_0 = H_{\mathrm{L}} + H_{\mathrm{C}} + H_{\mathrm{R}}$, defined as in \eqref{eq:green4:HL}--\eqref{eq:green4:HR}.
As before, we can use a Taylor expansion, to obtain
\begin{eqnarray}
 G_0(n,m) = \sum_{l=0}^{+\infty}\bra{n} \frac{\ccpar{H_{\mathrm{L}}+H_{\mathrm{C}}+H_{\mathrm{R}}}^l}{\ccpar{E+i\epsilon}^{l+1}}\ket{m}.
 \label{eq:G04:greentaylor}
\end{eqnarray}

Due to the same argumentation presented for the resonant level case, the division of the Hamiltonian in
three uncoupled terms allows us to write these elements as
\begin{small}
\begin{numcases}{G_0(n,m) =}
 \sum_{l=0}^{+\infty} \frac{\braketop{n}{H_{\mathrm{L}}^l}{m}}{\ccpar{E+i\epsilon}^{l+1}} = \bra{n} \frac{1}{E-H_{\mathrm{L}}+i\epsilon} \ket{m} \equiv \bra{n} G_{\mathrm{L}} \ket{m}, & $n,m<0$, \\
 \sum_{l=0}^{+\infty} \frac{\braketop{n}{H_{\mathrm{C}}^l}{m}}{\ccpar{E+i\epsilon}^{l+1}} = \bra{n} \frac{1}{E-H_{\mathrm{C}}+i\epsilon} \ket{m} \equiv \bra{n} G_{\mathrm{C}} \ket{m}, & $n,m=b_{i,j}$, \\
 \sum_{l=0}^{+\infty} \frac{\braketop{n}{H_{\mathrm{R}}^l}{m}}{\ccpar{E+i\epsilon}^{l+1}} = \bra{n} \frac{1}{E-H_{\mathrm{R}}+i\epsilon} \ket{m} \equiv \bra{n} G_{\mathrm{R}} \ket{m}, & $n,m>0$.
\end{numcases}
\end{small}
\noindent where $i,j=1,\dots,6$. As before, we shall study these regions separately.

\subsection{Calculation for the 1D chains next to the defect}
\label{sec:G04:LR}

The 1D chains surrounding the benzene molecule are similar to the ones surrounding the defect,
in the resonant level system. Therefore, the matrix elements in these regions are equal to the 
ones calculated in the previous appendix, what means that the expression \eqref{eq:G0:chains}, is
still valid for the benzene system, equally for $n \times m >0$.

\subsection{Calculation for the benzene molecule}
\label{sec:G04:C}

In this region, we want to find the elements
\begin{equation}
 G_0(b_n,b_m) = G_{\mathrm{C}}(b_n,b_m) = \braketop{b_n}{G_{\mathrm{C}}}{b_m},
\end{equation} 
\noindent where
\begin{equation}
 G_{\mathrm{C}} = \frac{1}{E-H_{\mathrm{C}}+i\epsilon},
\end{equation} 
\noindent and the Hamiltonian $H_{\mathrm{C}}$ is defined in \eqref{eq:green4:HC}. However, unlike 
the previous case, the Hamiltonian $H_{\mathrm{C}}$ is not diagonal in a basis composed by the 
position states, so we must now find its eigenstates $\ket{\psi_{\mathrm{C}}}$. To do so,
we start by assuming that they have the form
\begin{equation}
 \ket{\psi_{\mathrm{C}}}=\sum_{j=1}^6 c_{\mathrm{b}}(j) \ket{b_j},
 \label{eq:green4:psiC0}
\end{equation} 
\noindent and that, now, the coefficients $c_{\mathrm{b}}(j)$ have the form
\begin{equation}
 c_{\mathrm{b}}(j) = c_{\mathrm{b}}(0) e^{i  k_{\mathrm{b}} a j},
\end{equation}
\noindent where $k_{\mathrm{b}}$ is the wave vector of the waves propagating in the molecule. At this point,
a similar procedure to the one carried out in the previous section can be carried out now, from 
where it follows that
\begin{equation}
 \ket{\psi_{\mathrm{C}}}= \frac{1}{\sqrt{6}}\sum_{j=1}^6 e^{i  k_{\mathrm{b}} a j} \ket{b_j},
 \label{eq:green4:psiC1}
\end{equation} 
\begin{equation}
 k_{\mathrm{b}} \equiv k_{\mathrm{b}}(l) = \frac{2 \pi l}{6 a},
  \label{eq:green4:kb}
\end{equation} 
with $l=1,\dots,6$, and also
\begin{equation}
  E(k_{\mathrm{b}}) \equiv E_{\mathrm{b}}(k_{\mathrm{b}}) = - \varepsilon_0 - 2 t_{\mathrm{b}} \cos \ccpar{k_{\mathrm{b}} a}.
  \label{eq:green4:Eb}
\end{equation}

\noindent Since the states $\ket{\psi_{\mathrm{C}}(l)} \equiv \ket{k_{\mathrm{b}}(l)}$ compose a complete basis,
we can write
\begin{equation}
 G_0(b_n,b_m) = \sum_{l=1}^6 \braket{b_n}{k_{\mathrm{b}}(l)}\braketop{k_{\mathrm{b}}(l)}{G_{\mathrm{C}}}{b_m}.
\end{equation} 

\noindent The advantage of doing so is that $H_{\mathrm{C}} \ket{k_{\mathrm{b}}(l)} = E_{\mathrm{b}}\ccpar{k_{\mathrm{b}}} \ket{k_{\mathrm{b}}(l)}$,
which means that (omitting the index $l$, for simplicity),
\begin{eqnarray}
 G_0(b_n,b_m) &= \sum_{l=1}^6 \braket{b_n}{k_{\mathrm{b}}} \frac{\bra{k_{\mathrm{b}}} \ccpar{H_{\mathrm{C}}}^j \ket{b_m}}{\ccpar{E+i\epsilon}^{j+1}} \\
              &= \sum_{l=1}^6 \frac{\braket{b_n}{k_{\mathrm{b}}}\braket{k_{\mathrm{b}}}{b_m}}{E- E_{\mathrm{b}}\ccpar{k_{\mathrm{b}}}+i\epsilon}.
\end{eqnarray} 

\noindent Since we already know how to calculate the elements $\braket{b_n}{k_{\mathrm{b}}}$ and $\braket{k_{\mathrm{b}}}{b_m}$,
and we also know the explicit form of $E_{\mathrm{b}}\ccpar{k_{\mathrm{b}}}$, we finally obtain
\begin{eqnarray}
 G_0(b_n,b_m) &= \frac{1}{6} \sum_{l=1}^6 \frac{e^{i k_{\mathrm{b}}(l) a n} e^{-i k_{\mathrm{b}}(l) a m}}{E+ \varepsilon_0 + 2 t_{\mathrm{b}} \cos \sqpar{k_{\mathrm{b}}(l) a}} \\
              &= \frac{1}{6} \sum_{l=1}^6 \frac{e^{2 \pi i l (n-m)/6} }{E+ \varepsilon_0 + 2 t_{\mathrm{b}} \cos \ccpar{ \frac{2 \pi l}{6} }}.
\end{eqnarray} 

In short, the matrix elements of $G_0$ for the benzene system are
\begin{numcases}{ G_0(n,m) =\label{eq:G04:G0nm}}
 \frac{i}{2t\sqrt{1-\ccpar{\frac{E}{2t}}^2}} \ccpar{e^{i k a \abs{n+m}} - e^{i k a \abs{n-m}}}, & $n \times m>0$, \\
 \frac{1}{6} \sum_{l=1}^6 \frac{e^{2 \pi i l (n-m)/6} }{E+ \varepsilon_0 + 2 t_{\mathrm{b}} \cos \ccpar{ \frac{2 \pi l}{6} }} & $n,m=b_{i,j}$, \\
 0, & otherwise.
\end{numcases} 
\noindent where $i,j=1,\dots,6$.

\section*{References}


\begin{thebibliography}{99}
\bibitem{footnote1}
This is due to the separation of $H_0$ in
three independent and uncoupled parts, and is further explored in \ref{sec:anexoG03}.
\bibitem{footnote2}
It is now only presented the study of the transmittance. The case of the reflectance in 
analogue, since $\mathcal{R}=1-\mathcal{T}$.
\end{thebibliography}
\end{document}